\documentclass[draft]{aipproc}

\layoutstyle{6x9}

\begin{document}

\title{Magnetic Phases in Dense Quark Matter}


\author{Vivian de la Incera}
{address={Department of Physics, Western Illinois University,
Macomb, IL 61455, USA}}

\begin{abstract}
In this paper I discuss the magnetic phases of the three-flavor
color superconductor. These phases can take place at different field
strengths in a highly dense quark system. Given that the best
natural candidates for the realization of color superconductivity
are the extremely dense cores of neutron stars, which typically have
very large magnetic fields, the magnetic phases here discussed could
have implications for the physics of these compact objects.
\end{abstract}

\maketitle


\section{Introduction}

One of the fundamental tasks of modern nuclear physics is to
understand the phase structure of QCD. The study of the phase
structure of QCD requires exploring regions of different baryon
densities and temperatures. As known, at very high temperatures QCD
should be in a deconfined phase: the quark-gluon plasma. This phase
existed in the early universe and might be earthly reproduced in
high energy experiments. A deconfined phase can also exist in a very
different region of the temperature/density plane, that of low
temperatures and very high densities. This region is often called
dense QCD, although it really means cold and dense QCD. In this
environment the baryons get so squeezed that they start to overlap,
erasing any vestige of structure and liberating the quarks inside
them.

The ground state of a superdense quark system, a Fermi liquid of
weakly interacting quarks, is unstable with respect to the formation
of diquark condensates, a non-perturbative phenomenon essentially
equivalent to the Cooper instability of BCS superconductivity. In
QCD, one gluon exchange between two quarks is attractive in the
color-antitriplet channel. Thus, at sufficiently high density and
sufficiently small temperature, quarks should condense into Cooper
pairs, which are color antitriplets. These color condensates break
the SU(3) color gauge symmetry of the ground state producing a color
superconductor (CS) \cite{CS}.

The most likely candidates for the realization of the phenomenon of
color superconductivity are the core of compact stars, whose density
can reach values several times larger than the saturation density of
nuclear matter. Besides being very dense objects, compact stars
typically have very large magnetic fields too. In the case of
magnetars \cite{magnetars}, fields in the range of $10^{14} -
10^{16}$G have been predicted and observed. The quest to find
observable signatures of a color-superconducting core must not
ignore the presence of the star's magnetic field and its effects in
the superconducting state.

Although a color superconductor is in principle an electric
superconductor, because the diquark condensate carries nonzero
electric charge, in the color-flavor-locked (CFL) phase
\cite{alf-raj-wil-99/537} (CFL is the color-superconducting phase
that is realized in a system of three massless flavors at high
densities) there is no Meissner effect for a new in-medium
electromagnetic field $\widetilde{A}_{\mu}$. This in-medium (also
called "rotated") electromagnetic field is a combination of the
regular electromagnetic field and the $8^{th}$ gluon
\cite{alf-raj-wil-99/537,Gorbar-2000}. As the quark pairs are all
neutral with respect to the "rotated" electromagnetic charge
$\widetilde{Q}$, the "rotated" electromagnetic field
$\widetilde{A}_{\mu}$ remains long-range within the superconductor.

Because of the lack of a Meissner effect in the color
superconductor, the last one can be penetrated by the rotated
component of an external magnetic field. Depending on its strength,
an external magnetic field can then produce two different phases on
a three-flavor color superconductor: the
Magnetic-Color-Flavor-Locked (MCFL) phase \cite{MCFL} and the
Paramagnetic CFL (PCFL) one \cite{Vortex,PCFL}. The properties and
relevance of each of them at different magnetic scales have been
investigated in Ref. \cite{Ferrer-Incera-0703034}. The present paper
summarizes the main characteristics of these phases, as well as the
region of field strengths and chemical potentials where they (and
the CFL) can occur, and the type of transition that takes from one
to another.

\section{CFL-MCFL transmutation}

At zero magnetic field, the ground state of a three-massless-flavor
system at very high density is the CFL phase
\cite{alf-raj-wil-99/537}. However, once a magnetic field is
switched on, the difference between the electric charge of the $u$
quark and that of the $d$ and $s$ quarks reduces the original flavor
symmetry of the theory \cite{miransky-shovkovy-02} and consequently
also the symmetry group remaining after the diquark condensate is
formed. This less-symmetric color-superconducting phase is the MCFL
phase \cite{MCFL}. In this phase the external magnetic field
modifies the structure and magnitude of the quark gap in such a way
that the pairing of (rotated) electrically charged quarks is
reinforced by the external field. At very strong fields -of the
order of the baryon chemical potential- the pairing reinforcement is
sufficient to produce a distinguishable splitting of the gap in two
pieces: one that only gets contributions from pairs of neutral
quarks and one that gets contributions from both pairs of neutral
and pairs of charged quarks.

The CFL and MCFL phases not only differ in the structure and
magnitude of the gaps. They also have different low-energy physics.
The formation of the diquark condensate in the CFL case is
accompanied by the appearance of nine Goldstone bosons: a singlet
associated to the breaking of the baryonic symmetry $U(1)_B$, and an
octet associated to the axial $SU(3)_A$ group. Four fields of the
octet are charged with respect to the rotated electric charge. On
the other hand, the symmetry breaking that gives rise to MCFL leaves
a smaller number of Goldstone fields, all of which are neutral with
respect to the rotated electric charge. This implies that no charged
low-energy excitation can be produced in the MCFL phase, an effect
with possible consequences for the low-energy physics of a
color-superconducting star's core.

Despite the difference between MCFL and CFL, the two phases are
hardly distinguishable at weak magnetic fields. The symmetry of the
CFL phase can be considered as an approximated symmetry in the
presence of an external magnetic field \cite{ManuelJHW-05}, as long
as the field-induced mass of the charged Goldstone bosons is smaller
than twice the gap, so these mesons cannot decay into a
quasiparticle-quasihole pair \cite{Ferrer-Incera-0703034}. Since,
strictly speaking, the exact symmetry in the presence of the
magnetic field (ignoring quark masses) is that of MCFL, the
transition from the "approximated" CFL to MCFL at some threshold
field is not a phase transition, but a crossover or symmetry
transmutation.

The threshold field at which CFL ceases to be a good, approximate
symmetry can be found from the CFL low-energy theory of the
Goldstone bosons in the presence of a weak ($k^2 \sim {\tilde
e}{\tilde B}\ll\mu^2$) constant magnetic field \cite{ManuelJHW-05}.
The leading-order Lagrangian was found in
\cite{Ferrer-Incera-0703034}. It is given by

\begin{eqnarray}\label{chiral-lag-2}
L =\int d^{4}x \frac{f_\pi^2}{4}\{
[\sum_{A=1}^{3,8}|\partial_{0}\phi^{A}|^2+|\partial_{0}\Upsilon|^2]+v_{\|}^2[\sum_{A=1}^{3,8}|\partial_{\|}\phi^{A}|^2+|\partial_{\|}\Upsilon|^2]
\nonumber \\
+v_{\bot}^2[\sum_{A=1}^{3,8}|\partial_{\bot}\phi^{A}|^2+|(\partial_{\bot}+i\widetilde{e}\widetilde{A}_{\bot})\Upsilon|^2]\},
\qquad \qquad \qquad  \qquad \qquad
\end{eqnarray}
where the fields $\phi^A$ denote the neutral mesons and
\begin{eqnarray}  \label{neutr-inv-propg}
\Upsilon \equiv \left(
\begin{array}{cc}
\Pi^{+}\\
\kappa^{+}
\end{array}
\right) \
\end{eqnarray}
represents the charged meson doublet \cite{Ferrer-Incera-0703034}.

To find the strength of the threshold field from
(\ref{chiral-lag-2}), one can go to momentum space by applying the
Ritus' method \cite{Ritus-method} to the case with charged scalar
fields \cite{Ferrer-Incera-0703034} in order to obtain the
dispersion relation of the charged mesons
\begin{equation}
E^2 = \widetilde{e}\widetilde{B}(2n+1)v_{\bot}^2+k_{3}^2 v_{\|}^2.
\end{equation}
Taking into account that at zero momentum $(k_{3}=0, n=0)$ the rest
energy of the charged mesons is $M_{\widetilde{B}}^2 =
\widetilde{e}\widetilde{B}v_{\bot}^2$, it is clear that these mesons
acquire a mass due to the nonzero magnetic field
\cite{Ferrer-Incera-0703034}. These bound states will decay when
their mass is equal or larger than their constituents, that is,
twice the gap. Hence, the threshold field is
\begin{equation}\label{Threshold-value}
\widetilde{e}\widetilde{B}_{MCFL}=\frac{4}{v_{\bot}^2}\Delta_{CFL}^2\simeq
12\Delta_{CFL}^2,
\end{equation}
where the velocity $v_{\bot}\approx 1/\sqrt{3}$  was approximated by
its zero-field value \cite{Stephanov}. Using $\Delta_{CFL} \sim 15
MeV$, the order of magnitude of the field strength at which the
charged mesons decouple is $\sim 10^{16}G$.

After this decoupling, the low-energy physics of the system is
driven by the five neutral Goldstone bosons (including the one
associated with the baryon-symmetry breaking) that characterize the
MCFL phase. Therefore, the MCFL phase becomes relevant at magnetic
field scales comparable to the CFL gap square, even though near the
threshold field the splitting of the gaps is still negligible.

Notice the analogy between the CFL-MCFL transmutation and what can
be called a "field-induced" Mott transition
\cite{Ferrer-Incera-0703034}. Mott transitions have been discussed
in condensed matter and in QCD \cite{Mott-1,Mott-2} to describe
delocalization of bound states into their constituents at a
temperature defined as the Mott temperature. By definition, the Mott
temperature $T_{M}$ is the temperature at which the mass of the
bound state equals the mass of its constituents, so the bound state
becomes a resonance at $T>T_{M}$. In the CFL-MCFL transmutation, the
usual role of the Mott temperature is played by the threshold field
$\widetilde{B}_{MCFL}$.

\section{MCFL-PCFL phase transition}

The magnetic field can also influence the gluon dynamics
\cite{Vortex,PCFL} because some of the gluons in the CFL have
nonzero rotated charge. As shown above, at field strengths larger
than $\widetilde{B}_{MCFL}$, the system is in the MCFL phase. For
fields in the range: $\widetilde{e}\widetilde{B}_{MCFL}\leq
\widetilde{e}\widetilde{B}< m^{2}_{M}$, where the Meissner mass
square $m^{2}_{M}$ of the charged gluons satisfies $\mu^{2} \gg
m^{2}_{M}\gg \Delta^{2}_{CFL}$, the field-induced gap separation is
still negligible, as corroborated by recent numerical calculations
\cite{numerical-MCFL}. That allows one to assume that the quark gap
is basically given by $\Delta_{CFL}$, even at field strengths
comparable to $m^{2}_{M}$. This was the approach followed in papers
\cite{Vortex,PCFL}. It led to the discovery of a new magnetic phase:
the PCFL one. Let us explain how it appears.

Once $\widetilde{B} \geq \widetilde{B}_{PCFL} = m_{M}^2$, one of the
modes of the charged gauge field becomes tachyonic (this is the well
known "zero-mode problem" for spin-1 charged fields in the presence
of a magnetic field found for Yang-Mills fields \cite{zero-mode},
for the $W^{\pm}_{\mu}$ bosons in the electroweak theory
\cite{Skalozub, Olesen}, and even for higher-spin fields in the
context of string theory \cite{porrati}). Similarly to other spin-1
theories with magnetic instabilities \cite{zero-mode}-\cite{Olesen},
the solution of the zero-mode problem leads to the restructuring of
the ground state through the formation of an inhomogeneous
gauge-field condensate $G$, as well as an induced magnetic field due
to the back reaction of the $G$ condensate on the rotated
electromagnetic field. The magnitude of the $G$-condensate plays the
role of the order parameter for the phase transition occurring at
$\widetilde{B}= \widetilde{B}_{PCFL}$.

Near the transition point, the amplitude of the condensate $G$ is
very small \cite{Vortex}. Then, the condensate solution can be found
using a Ginzburg-Landau (GL) approach similar to Abrikosov's
treatment of type II metal superconductivity near the critical field
$H_{c2}$ \cite{Abrikosov}. As in Abrikosov's case, the order
parameter $|G|$ continuously increases from zero with the applied
magnetic field, signalizing a second-order phase transition towards
a gluon crystalline vortex state characterized by the formation of
flux tubes. Both spatial symmetries -the rotational symmetry in the
plane perpendicular to the applied magnetic field and the
translational symmetry- are broken by the vortex state.

Nevertheless, contrary to what occurs in conventional type-II
superconductors, where the applied magnetic field only penetrates
through flux tubes and with a smaller strength than that of the
applied field, the gluon vortex state exhibits a paramagnetic
behavior. That is, outside the flux tube the applied field
$\widetilde{B}$ totally penetrates the sample, while inside the
tubes the magnetic field becomes larger than $\widetilde{B}$. This
antiscreening behavior is similar to that found in the electroweak
system at high magnetic field \cite{Olesen}. Hence, since the
$\widetilde{Q}$ photons remain long-range in the presence of the
condensate $G$, the rotated electromagnetism remains unbroken. At
asymptotically large densities, because $\Delta_{CFL}\ll m_{M}$, we
have $\widetilde{B}_{MCFL}\ll \widetilde{B}_{PCFL}$ for each $\mu$
value.

\section{Magnetic Phases}\label{MP}

\begin{figure}
\includegraphics[height=.3\textheight]{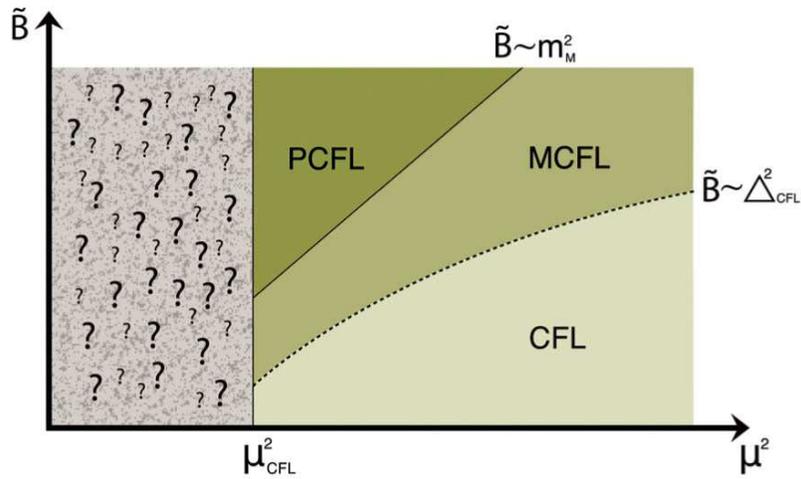}
\caption{Qualitative sketch in the $\widetilde{B}$ vs $\mu^2$ plane
of the different phases of a color superconductor with three quark
flavors in the presence of an external magnetic field at
asymptotically high densities. The CFL phase appears here as an
approximate symmetry at weak field. Thus, the line between the CFL
and MCFL phases does not denote a real phase transition, but the
boundary separating the approximated CFL phase from the MCFL phase.
This symmetry-transmutation line is reached at field values of the
order of the CFL gap square. The line between the MCFL and PCFL
phases indicates a second-order phase transition curve occurring at
field strengths of the order of the magnetic mass square of the
charged gluons. The rectangular region to the left corresponds to
moderately high densities in the presence of a magnetic field. Since
the ground state at moderately high densities has not yet been
investigated in the presence of a magnetic field, this region is
indicated by question marks.} \label{fig1}
\end{figure}

In summary, at very high densities a three-flavor CS can be in one
of three phases, depending on the magnitude of the external magnetic
field. Going from low to high fields, a symmetry transmutation from
CFL to MCFL will take place first, and then a second-order phase
transition from MCFL to PCFL (see Fig.\ref{fig1}). During the
CFL-MCFL transmutation no symmetry breaking occurs, since in
principle once a magnetic field is present, the symmetry is
theoretically that of the MCFL. In practice, however, when
$\widetilde{B}< \widetilde{B}_{MCFL}\sim \Delta^{2}_{CFL}$, the MCFL
phase is almost indistinguishable from the CFL. Only at fields
comparable to $\Delta^{2}_{CFL}$ the main features of MCFL emerge
through the low-energy behavior of the system. At the threshold
field $\widetilde{B}_{MCFL}$, only five of the original nine
Goldstone bosons that characterize the low-energy behavior of the
CFL phase remain. These are precisely the five Goldstone bosons
determining the new low-energy behavior of the genuinely realized
MCFL phase. Going from MCFL to PCFL is, on the other hand, a real
phase transition \cite{Vortex,PCFL}, as the translational symmetry,
as well as the remaining rotational symmetry in the plane
perpendicular to the applied magnetic field, are broken by the
vortex state.

In the above considerations we have ignored the effects due to
non-zero quark masses, because we assumed a very large baryon
density. However, the densities of interest for most astrophysical
applications are just moderately high and for them the effects of
quark masses and color and electric neutrality should not be
ignored. It is plausible that if compact stars are the natural
playground for color superconductivity, the magnetic phases
described in this paper, or more precisely, the version of these
phases at more realistic densities, may be relevant for the physics
of the core of highly magnetized compact objects like magnetars. We
refer the interested reader to Refs. \cite{magnetar-criticism} for a
discussion of the inconsistencies between some observations and the
standard magnetar model, and to Ref.\cite{Ferrer-Incera-0703034} for
a discussion of the possible implications of the CS magnetic phases
in the solution of this conundrum.

\begin{theacknowledgments}
  This work has been supported in part by DOE Nuclear Theory
grant DE-FG02-07ER41458.
\end{theacknowledgments}



\bibliographystyle{aipproc}   



\end{document}